\documentclass[aps,prd,reprint,twocolumn,superscriptaddress,floatfix,nofootinbib,longbibliography]{revtex4-2}
\usepackage{float}
\usepackage{microtype}
\usepackage{graphicx, amsfonts, amsthm, xr}
\usepackage{array}
\usepackage{epsfig,amsmath, amssymb, color, dsfont, upgreek, physics}
\usepackage{mathrsfs}
\usepackage{mathtools}
\usepackage{bbold}
\usepackage{comment}
\usepackage{braket}
\usepackage{bm}

\usepackage[bookmarks=true,colorlinks,linkcolor=OrangeRed,urlcolor=NavyBlue,citecolor=RoyalBlue]{hyperref}
\usepackage[capitalise]{cleveref}
\Crefname{figure}{Fig.}{Figs.}
\Crefname{section}{Sec.}{Secs.}
\usepackage[dvipsnames]{xcolor}
\usepackage[inline]{enumitem}
\setlist[enumerate]{label=(\roman*)}
\usepackage{parskip}
\usepackage{xparse}
\usepackage{etoolbox}
\DeclareMathOperator{\diff}{d}
\DeclareMathOperator{\Ham}{\hat{H}}
\newcommand{\Norm}[1]{\left\lVert #1 \right\rVert} 
\newcommand{\rdens}[1][]{\varrho_{#1}}

\NewDocumentCommand{\field}{o o o}{%
  \hat{\varphi}%
  \IfValueT{#1}{_{#1}}%
  \IfValueT{#2}{%
    \IfValueTF{#3}
      {\ifstrequal{#3}{1}{^{#2\dagger}}{^{#2}}}
      {^{#2}}%
  }%
  \IfNoValueTF{#2}{%
    \IfValueT{#3}{%
      \ifstrequal{#3}{1}{^{\dagger}}{}%
    }%
  }{}%
}

\NewDocumentCommand{\mom}{o o o}{%
  \hat{\pi}%
  \IfValueT{#1}{_{#1}}%
  \IfValueT{#2}{%
    \IfValueTF{#3}
      {\ifstrequal{#3}{1}{^{#2\dagger}}{^{#2}}}
      {^{#2}}%
  }%
  \IfNoValueTF{#2}{%
    \IfValueT{#3}{%
      \ifstrequal{#3}{1}{^{\dagger}}{}%
    }%
  }{}%
}

\NewDocumentCommand{\Op}{m o o}{%
  \hat{#1}%
  \IfValueT{#2}{_{#2}}%
  \IfValueT{#3}{%
    \ifstrequal{#3}{1}{^{\dagger}}{}%
  }%
}
\newcommand{\phdag}{\vphantom{dagger}}
\NewDocumentCommand{\lattf}{smmm}{%
    \hat{#2}%
    _{\vb{#3}}%
    ^{\IfBooleanTF{#1}{\dagger\,}{\phdag}#4}%
}
\NewDocumentCommand{\matt}{sG{x}O{\alpha}}{\IfBooleanTF{#1}{\lattf*}{\lattf}{\psi}{#2,#3}{}}
\NewDocumentCommand{\gaugf}{smmmm}{\IfBooleanTF{#1}{\lattf*}{\lattf}{#2}{#3,\vb*{#4}}{#5}}
\NewDocumentCommand{\para}{sG{x}D<>{\mu}O{\alpha\beta}}{\IfBooleanTF{#1}{\gaugf*}{\gaugf}{U}{#2}{#3}{#4}}

\newcommand{\idest}{i.e.}

\newcommand{\hc}{{\rm{h.c}}}

\newcommand{\spin}{j}
\newcommand{\jmax}{\spin_{\max}}
\newcommand{\jonehalf}{\hspace{-0.7pt}{\frac12}\hspace{-0.7pt}}

\NewDocumentCommand{\casimireig}{O{\gtl}}{\mathcal{E}^2_{#1}}
\newcommand{\hpsi}{\hat{\psi}}
\newcommand{\vecsite}{\vb{n}}
\newcommand{\site}[1][]{n_{#1}}
\newcommand{\latvec}[1][]{\vb*{\mu}_{#1}}
\newcommand{\genlink}{{\vecsite, \latvec}}

\usepackage{orcidlink}
\newcommand{\orcidpeter}{\orcidlink{0009-0009-7020-7246}}
\newcommand{\orcidpietro}{\orcidlink{0000-0001-5279-7064}}
\newcommand{\orcidsimone}{\orcidlink{0000-0002-8882-2169}}
\newcommand{\orcidgiovanni}{\orcidlink{0000-0002-9073-8978}}
\newcommand{\DFA}{\affiliation{Dipartimento di Fisica e Astronomia ``G. Galilei'', Università di Padova, I-35131 Padova, Italy.}}
\newcommand{\PQTC}{\affiliation{Padua Quantum Technologies Research Center, Università degli Studi di Padova}}
\newcommand{\INFNPD}{\affiliation{Istituto Nazionale di Fisica Nucleare (INFN), Sezione di Padova, I-35131 Padova, Italy.}}
\newcommand{\MPQ}{\affiliation{Max Planck Institute of Quantum Optics, 85748 Garching, Germany}}
\newcommand{\MCQST}{\affiliation{Munich Center for Quantum Science and Technology (MCQST), 80799 Munich, Germany}}

\graphicspath{{figures/}}

\begin{document}

\title{Optimal local basis truncation of lattice quantum many-body systems}
\author{Peter Majcen$^{\orcidpeter}$}\DFA \PQTC \INFNPD
\author{Giovanni Cataldi$^{\orcidgiovanni}$} \DFA \PQTC \INFNPD \MPQ \MCQST
\author{Pietro Silvi$^{\orcidpietro}$} \DFA \PQTC\INFNPD 
\author{Simone Montangero$^{\orcidsimone}$} \DFA \PQTC \INFNPD 
\date{\today}

\begin{abstract}
We show how to optimally reduce the local Hilbert basis of lattice quantum many-body (QMB) Hamiltonians.
The basis truncation exploits the most relevant eigenvalues of the estimated single-site reduced density matrix (RDM). 
It is accurate and numerically stable across different model phases, even close to quantum phase transitions. 
We apply this procedure to different models, such as the Sine-Gordon model, the $\field^{4}$ theory, and lattice gauge theories, namely Abelian $\mathrm{U}(1)$ and non-Abelian $\mathrm{SU}(2)$, in one and two spatial dimensions.
Our results reduce state-of-the-art estimates of computational resources for  classical and quantum simulations.
\end{abstract}

\maketitle
In the study of QMB systems, the dimension $d$ of the single-site local Hilbert space plays a crucial role in determining the computational cost and feasibility of both classical and quantum simulations. 
Although paradigmatic spin models such as the Ising or Heisenberg Hamiltonians can operate with a minimal local basis $d\!=\!2$, most physically relevant systems exhibit large or unbounded local spaces arising from different physical mechanisms.
In these scenarios, basis truncation strategies become crucial for enabling numerical and experimental simulations.

From a computational perspective, an increase in the local dimension directly affects memory requirements and algorithmic complexity \cite{Eisert2013EntanglementTensorNetwork-1}. 
In the realm of tensor network (TN) algorithms \cite{Orus2014PracticalIntroductionTensor-1, Montangero2018IntroductionTensorNetwork-1}, such as Matrix Product States (MPS) \cite{Schuch2008EntropyScalingSimulability-1, Schollwock2011DensitymatrixRenormalizationGroup-1, Paeckel2019TimeevolutionMethodsMatrixproduct-1}, Projected Entangled Pair States (PEPS) \cite{Verstraete2006CriticalityAreaLaw-1, Schuch2007ComputationalComplexityProjected-1, Verstraete2008MatrixProductStates-1, Corboz2018FiniteCorrelationLength-1, Cirac2021MatrixProductStates-1, Vanderstraeten2022VariationalMethodsContracting-1}, or Tree Tensor Networks (TTN) \cite{Silvi2019TensorNetworksAnthology-1}, the cost of tensor storage and contraction scales with the local dimension, often quadratically or worse.
These challenges are not limited to classical simulations. 
On quantum hardware, while qudit-based quantum processors have significantly improved in gate fidelity and scalability \cite{Ringbauer2022UniversalQuditQuantum-1}, a large local dimension $d$ would require as many transitions between the internal levels, which are hard to engineer and not always available.
Alternatively, encoding large local Hilbert spaces into qubits substantially increases the computational cost by extending interaction ranges and introducing genuine many-body terms into the Hamiltonian~\cite{Setia2019SuperfastEncodingsFermionic-1}, which results in deeper circuit depths.

This work proposes a general and adaptive truncation scheme based on single-site RDM estimation, which naturally identifies the most relevant local states from the QMB ground-state or low-lying eigenstates. 
By construction, the resulting local basis is optimal under local transformations. 
This provides a fundamental pre-processing step for efficient classical or quantum simulations. 

\begin{figure}
\centering
\includegraphics[width=1\columnwidth]{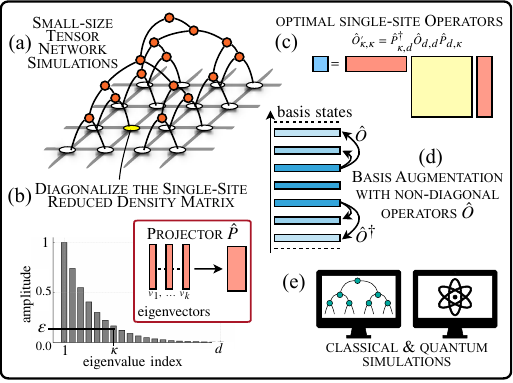}
\caption{
\textbf{Single-site optimal basis truncation protocol.}
(a) From small lattice-size -- exact diagonalization (ED), tensor networks (TN), mean-field (MF) or cluster-MF -- simulations of the given Hamiltonian, (b) extract and diagonalize the single-site RDM and build a projector $\hat{P}$ selecting the first $\kappa$ eigenstates $v_{1}\!\dots\!v_{k}$ corresponding to the eigenvalues $\lambda_{1}\!\dots\!\lambda_{k}$ larger than a certain numerical precision $\varepsilon$. 
(c) Project all the Hamiltonian operators on the new basis with $\hat{P}$. 
(d) The basis can be eventually augmented by applying non-diagonal local operators $\Op{O}$ of the theory on the existing projector $\hat{P}$. 
Finally, (e) the resulting optimal basis can be used for classical and quantum simulations.}
\label{fig1_scheme}
\end{figure}

Large local Hilbert spaces appear in a wide range of physical contexts.  
In bosonic lattice models, such as the Bose-Hubbard model, local site occupations are in principle unbounded due to bosonic statistics. 
However, strong interactions and low fillings, when working with finite energy bandwidths, naturally suppress high-occupancy states \cite{Hanada2023EstimatingTruncationEffects}, making truncated local Fock spaces both physically justified and commonly used in both exact diagonalization (ED) and density matrix renormalization group (DMRG) approaches \cite{Kuhner1999DynamicalCorrelationFunctions, Jaksch1998ColdBosonicAtoms}.
Important exceptions arise in systems with infinitely many states at finite bandwidth, such as electronic bound states in atoms.
Similarly, lattice gauge theories (LGT) \cite{Kogut1975HamiltonianFormulationWilsons-1, Rothe2012LatticeGaugeTheories-1} with continuous gauge groups such as $\mathrm{U(1)}$, $\mathrm{SU(2)}$, or $\mathrm{SU(3)}$ require infinite-dimensional gauge link Hilbert spaces. 
Truncated formulations, such as quantum link models (QLM) \cite{Chandrasekharan1997QuantumLinkModels-1, Horn1981FiniteMatrixModels-1, Orland1990LatticeGaugeMagnets-1, Brower1999QCDQuantumLink-1}, Casimir-energy cutoffs \cite{Cataldi2024Simulating2+1DSU2-1, Calajo2024DigitalQuantumSimulation-1, Rigobello2021EntanglementGeneration$1+1mathrmD$-1, Rigobello2023Hadrons1+1DHamiltonian-1, Calajo2025QuantumManybodyScarring, Cataldi2025DisorderFreeLocalizationFragmentation, Magnifico2025TensorNetworksLattice-1, Cataldi2025RealTimeStringDynamics-1}, 
finite subgroups \cite{Ercolessi2018PhaseTransitions$Z_n$-1, Magnifico2020RealTimeDynamics-1, Haase2021ResourceEfficientApproach-1, Mariani2023HamiltoniansGaugeinvariantHilbert-1}, digitization of gauge fields \cite{Hackett2019DigitizingGaugeFields-1}, and fusion-algebra deformation \cite{Zache2023QuantumClassicalSpinNetwork-1}, are essential for numerical treatment \cite{Banuls2017EfficientBasisFormulation-1, Magnifico2025TensorNetworksLattice-1} and experimental realization on quantum platforms \cite{Byrnes2006SimulatingLatticeGauge-1, Martinez2016RealtimeDynamicsLattice-1, Banuls2020SimulatingLatticeGauge-1, Mathis2020ScalableSimulationsLattice-1, Davoudi2020AnalogQuantumSimulations-1, Mazzola2021GaugeinvariantQuantumCircuits-1, Kan2021Investigating$3+1mathrmD$Topological-1, Zohar2021QuantumSimulationLattice-1, Pomarico2023DynamicalQuantumPhase-1, Bauer2023QuantumSimulationFundamental-1, Bauer2023QuantumSimulationHighEnergy-1}. 
The reliability of such truncations is often justified in strong coupling or low-energy limits \cite{Buyens2017FiniterepresentationApproximationLattice-1}.
Large local spaces also arise in electron-phonon systems~\cite{Jeckelmann1998DensitymatrixRenormalizationgroupStudy-1, Macridin2018ElectronPhononSystemsUniversal}, such as the Holstein and SSH models, which involve fermions coupled to local harmonic oscillators, resulting in an infinite local basis. 
Tractable subspaces are obtained via phonon cutoffs based on coupling strength and occupation statistics~\cite{Bonca1999HolsteinPolaron, Franchini2021PolaronsMaterials}. 
A similar task is faced in multi-orbital, impurity, and molecular systems due to spin, orbital, and charge configurations \cite{Georges1996DynamicalMeanfieldTheory}, as well as in quantum chemistry from vibrational and rotational modes requiring energy-based truncation or mode reduction \cite{Baiardi2020DensityMatrixRenormalization}.

Despite their differences, all these systems highlight a common challenge: identifying and retaining the physically relevant subspace of a large or infinite local Hilbert space. 
While energy cutoffs, symmetry constraints, or system-specific intuition are often used, all these truncation strategies are typically model-dependent and may fail near criticality or in strongly entangled regimes.

Existing local basis optimization (LBO) methods are typically described as techniques that adaptively truncate or transform the large local Hilbert space into a smaller, more relevant effective basis during DMRG sweeps~\cite{Stolpp2021ComparativeStudyStateoftheart-1, Stolpp2020ChargedensitywaveMeltingOnedimensional-1, Guo2012CriticalStrongCouplingPhases-1}. 
In contrast, this work proposes a different strategy: a pre-processing step in which the RDM is estimated using both ED, MF, and TN calculations. 
This makes our approach independent of the specific problem or TN architecture, and directly applicable to quantum simulation.

The paper is organized as follows: in \cref{sec_LBO_scheme}, we introduce the numerical scheme for local basis optimization and present an algorithm for basis augmentation.
In \cref{sec_scalar_field_theories}, we apply the method to scalar field theories in $(1\!+\!1)\mathrm{D}$ and $(2\!+\!1)\mathrm{D}$, such as the Sine-Gordon model and the $\field^4$-theory, and identify the regimes where basis truncation provides the greatest advantage. 
In \cref{sec_basis_augmentation}, we illustrate how basis augmentation can improve the local Hilbert space expansion in terms of the RDM eigenvectors. 
We then extend our analysis to $\mathrm{U(1)}$ and $\mathrm{SU(2)}$ LGTs in \cref{sec_LGTs}, comparing the conventional electric basis with the one obtained through our optimization scheme. 
Finally, we discuss in \cref{sec_quantum_simulation} the implications of the optimal basis reduction scheme for TN and digital quantum simulations.

\section{Local basis optimization scheme}
\label{sec_LBO_scheme}
The fundamental steps towards an optimal truncation of a large local basis for a generic QMB system are sketched in \cref{fig1_scheme}.
We consider a QMB theory defined on a $D$-dimensional spatial lattice $\Lambda$, with $N$ sites $\vecsite\!=\!(\site[1],\dots\site[D])$ equally spaced by lattice spacing $a>0$ and links identified by positive unit vectors $\latvec[k]$ with $k\in\qty{1\!\dots\!D}$. 
Assuming the theory degrees of freedom to be originally described with a local Hilbert space of size $d=\dim\mathcal{H}_{\vecsite}$, there might exist (in principle) a smaller and numerically exact (up to a certain precision) basis which can be obtained (\emph{a posteriori}) by computing the single-site RDM.
In practice, if $d$ is significantly large, any classical or quantum simulation of the problem would be limited to small lattice sizes, preventing access to such an exact basis. 
We then need to find an \emph{optimal} basis out of the available resources we can access.

\textbf{\textit{(\romannumeral 1) RDM estimation.---}} For a fixed value of the Hamiltonian parameters, we initially focus on the smallest non-trivial lattice size where we can isolate a single site in the bulk of a translational invariant portion of the system (e.g., three sites in a nearest-neighbor 1D system).
To isolate a single site, we exploit three approaches: (\emph{i}) by using a $k$-site cluster mean-field (MF) approach; (\emph{ii}) through ground-state TN simulations with maximal bond dimension $\chi$ on the smallest available lattice size; (\emph{iii}) with ED methods on the largest available translational invariant lattice. 
Once we access the ground state estimate of the problem, we compute the RDM $\rdens[i]$ of the $i^{\rm{th}}$ site in the bulk of the system. 
If the wave function ansatz lacks translational invariance, as in the case of small-size TTN calculations, an efficient basis can still be extracted by computing the mean RDM over the system size.
By diagonalizing $\rdens[i]$, we then extract and sort in descending order the corresponding $d$ eigenvalues $\qty{\lambda_{1}\dots \lambda_{d}}$ and the associated $d$-dimensional eigenstates $\qty{\vb{v}_{1}\dots \vb{v}_{d}}$. 

\textbf{\textit{(\romannumeral 2 ) Basis truncation.---}} Then, we set a finite and controllable tolerance $\varepsilon$ which selects the first $k$ eigenstates $\qty{\vb{v}_{1}{\dots} \vb{v}_{k<d}}$ such that $\lambda_{j}{\geq} \varepsilon$, $\forall j=1{\dots}k\!<\!d$. 
Out of these eigenstates, we build a projector $\Op{P}[d,k]$ which projects the single-site Hamiltonian operators $\Op{O}$ in the Hilbert subspace: $\Op{O}^{\prime}_{k,k}\!=\!\Op{P}[k,d][1]\Op{O}_{d,d}\Op{P}[d,k]$. 
The optimal Hamiltonian obtained using the truncated operators $\Op{O}^{\prime}_{k,k}$ allows for simulating the original model with a smaller local Hilbert space and exploring larger system sizes.

\begin{figure*}[ht]
    \centering
    \includegraphics[width=1\textwidth]{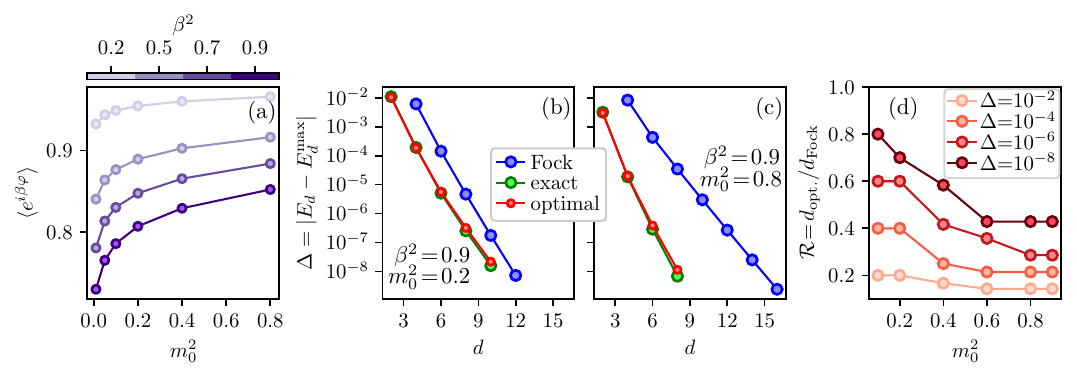}
\caption{\textbf{Local basis optimization of the Sine-Gordon theory.}
(a) Expectation value of the vertex operator $\braket{e^{i\beta \field}}$ as a function of the bare mass $m_0^2$, for fixed values of the coupling $\beta^2$ obtained from MPS simulation on a $32$-sites chain with bond-dimension $\chi\!=\!64$. 
(b–c) Convergence of the ground state energy prediction $E_{d}$ as a function of the local Hilbert space dimension $d$ using the Fock basis (blue), the exact basis (green), and the optimal basis (red) derived from $3$-site MF calculations. 
The reference values $E_{d}^{\max}$ are obtained from MPS simulations on a $32$-sites chain with bond-dimension $\chi\!=\!64$ using the Fock basis with $d_{\max}=16$ (b) and $d_{\max}=18$ (c) respectively.
(d) Ratio between the local dimensions of the optimal and the original Fock bases, corresponding to some target accuracies $\Delta\in \{10^{-2},10^{-4},10^{-6},10^{-8}\}$ in the ground-state energy, as a function of the bare mass $m_{0}^2$.}
\label{fig2_sinegordon}
\end{figure*}
\textbf{\textit{(\romannumeral 3 ) Basis augmentation.---}} Such a procedure strongly relies on the scaling of eigenvalues in the single-site RDM, which in turn varies across different regimes of the zero-temperature phase diagram depending on the distance to any critical point of the system. 
In their proximity, some long-range modes might not appear in the single-site $\rdens[i]$ obtained from too-small-lattice simulations. 
To verify the reliability of this LBO protocol, especially when expanding the local Hilbert space into the $\rdens[i]$-basis fails to yield a good ground-state approximation, we propose an \emph{augmentation} scheme (see \cref{fig1_scheme}(e)).
We select powers of single-site non-diagonal Hamiltonian operators $\Op{O}[i]$ and iteratively apply each of them to the $k$ most relevant eigenstates $\qty{\vb{v}_{1}{\dots} \vb{v}_{k}}$ of $\rdens[i]$, 
\begin{equation}
 \Op{O}_{i}\qty(\vb{v}_{1},\dots,\vb{v}_{k})=
 \qty(\vb{v}_{1}^{\prime},\dots ,\vb{v}_{k}^{\prime}),
\end{equation}
searching for other relevant states with a small overlap with the selected largest $k$ ones. 
Specifically, for a chosen residual tolerance $r>0$ such that
\begin{equation}
    \Norm{\vb{v}_{i}^{\prime}-\sum_{\ell=1}^{k} \braket{\vb{v}_{i}^{\prime}|{\vb{v}_{\ell}}  }\vb{v}_{\ell}}_{2} >r, 
    \label{eq_tolerance_augmentation}
\end{equation}
we extend the existing basis $\qty{\vb{v}_{1}\dots \vb{v}_{k}}$ to $\qty{\vb{v}_{1}\dots \vb{v}_{k}, \vb{v}_{i}^{\prime}}$ and apply the Gramm-Schmidt algorithm to it to obtain an orthonormal spanning set. 
If none of the obtained states is found to be orthonormal to the eigenstates of $\rdens[i]$, the truncation scheme is already optimal; otherwise, the projector $\Op{P}$ is augmented from $k$ to $k +\delta k$ with $\delta k$ new states satisfying \cref{eq_tolerance_augmentation}.

\section{Scalar Field Theories}
\label{sec_scalar_field_theories}
As a first application, we implement the protocol on two paradigmatic scalar field theories: the sine-Gordon model \cite{Lukyanov1997ExactExpectationValues, ZAMOLODCHIKOV1995MASSSCALESINE, KLASSEN1993SINEGORDONVSMASSIVE, Calliari2024QuantumSimulatingContinuum} and the $\field^4-$ model \cite{Loinaz1998MonteCarloSimulation, Sugihara2004DensityMatrixRenormalization, Milsted2013MatrixProductStates-1}. 
Throughout the work, we work in the units where $\hbar\!=\!c\!=\!1$. 
To obtain the corresponding discretized Hamiltonians, the $D$-spatial dimensions are regularized (see \cref{app_scalarmodels}) on the lattice $\Lambda$. 
The obtained lattice field operator $\field_{\vecsite}$ and its conjugate momentum $\mom[\vecsite]=a\, \partial_{t}\field[\vecsite]$ satisfy the equal-time commutation relation: 
\begin{equation}
    [\field[\vecsite](t),\mom[\vecsite^{\prime}]{(t)}]=i\delta_{\vecsite,\vecsite^{\prime}} \qquad \forall \vecsite,\vecsite^{\prime}\in \Lambda
\end{equation}  
and can be rewritten in terms of real-space realizations of creation and annihilation operators $\Op{b}[\vecsite][1], \Op{b}[\vecsite]$, 
\begin{align}
    \field[\vecsite]&=\frac{1}{\sqrt{2}}\left(\Op{b}[\vecsite][1]+\Op{b}[\vecsite]\right)&
    \mom[\vecsite]&=\frac{i}{\sqrt{2}}\left(\Op{b}[\vecsite][1]-\Op{b}[\vecsite]\right),
\end{align}
with $\Op{b}[\vecsite] \ket{0}=0$, and $\Op{b}[\vecsite]$, $\Op{b}[\vecsite][1]$ fulfilling the canonical commutation relation $[\Op{b}_{\vecsite},\Op{b}[\vecsite^{\prime}]^{\dagger}]=\delta_{\vecsite,\vecsite^{\prime}}$ whereas all the other commutation relations are zero.

\subsection{Sine-Gordon model}
\label{sec_sG-model}
The sine-Gordon (sG) model in $(1\!+\!1)\mathrm{D}$ is a well-known example of an integrable relativistic quantum field theory.
Due to its integrability, the model represents an ideal test ground for our LBO protocol.
The corresponding lattice Hamiltonian (in units of the lattice spacing $a=1$) reads: 
\begin{equation}
    \Ham\!=\!\sum_{\vecsite\in\Lambda}\qty[\frac{\mom[\vecsite][2]}{2}\!+\!\frac{1}{2}\qty[\qty(\field[\vecsite+\latvec]\!-\!\field[\vecsite])^2\!-\!\frac{m_{0}^{2}}{\beta^{2}}\cos(\beta\field[\vecsite])]],
    \label{eq_sinegordon_hamiltonian}
\end{equation}
The dimensionless coupling $\beta^{2}\!>\!0$ identifies the different phases of the models \cite{Lukyanov1997ExactExpectationValues}. 
Namely, at $\beta^2\!=\!8\pi$, the model exhibits a Berezinsky-Kosterlitz-Thouless (BKT) transition, which separates a gapless (and renormalizable) Luttinger liquid phase in $\beta^2\!>\!8\pi$ from a gapped phase in $\beta^2\!<\!8\pi$ supporting massive single-particle excitations. 
The continuous theory exhibits no divergences in the regime of $\beta^2\!<\!4\pi$, while being super-renormalizable in $4\pi\!<\!\beta^2<8\pi$.
The model has a $\mathbb{Z}_{n}$ symmetry $\field \rightarrow \field\!+\!2\mom n/\beta$, where $n$ is an arbitrary integer. 

Throughout this work, we study the model in the parameter regime $0\!<\!\beta\!<\!1$, where the symmetry $\mathbb{Z}_{n}$ is spontaneously broken. 
In this \emph{broken} phase, the theory has infinitely many groundstates $\ket{0_{n}}$, characterized by a finite, nonzero expectation value of the order parameter, which in the sG theory corresponds to the vertex operator and is analytically known \cite{Lukyanov1997ExactExpectationValues} and reads:
\begin{equation}
\begin{split}
    \langle e^{i\beta \field}\rangle=&
    \frac{(1+\xi)\pi \Gamma (\frac{1}{1+\xi})}{16 \sin(\pi \xi) \Gamma(\frac{\xi}{1+\xi})}\qty[\frac{\Gamma(\frac{1+\xi}{2})\Gamma(1-\frac{\xi}{2})}{4\sqrt{\pi}}]^{-\frac{2}{1+\xi}} m^{\frac{2\xi}{1+\xi}},
    \label{eq:sG-order_parameter}
\end{split}
\end{equation}
where \begin{math}
    \xi\!=\!\beta^2 /(8\pi\!-\!\beta^2),
\end{math}
while $m\!=\!2M\sin(\pi\xi/2)$ is the mass of the lightest bound state of the particle associated with $\field$, expressed in terms of the soliton mass $M$, which is analytically known \cite{Lukyanov1997ExactExpectationValues, Calliari2024QuantumSimulatingContinuum} (see \cref{eq:sG-soliton_mass} in \cref{sec_appendix-sG}).

The need for a large local Hilbert space becomes particularly significant in the \emph{broken} phase of the theory (see \cref{fig2_sinegordon}(a)), where higher field excitations are required to mitigate truncation effects caused by the growth of the order parameter $\braket{e^{i\beta \field}}$. 
Close to criticality, additional excitations are needed to capture large local fluctuations.

\begin{figure*}[!t]
    \centering
    \includegraphics[width=1\textwidth]{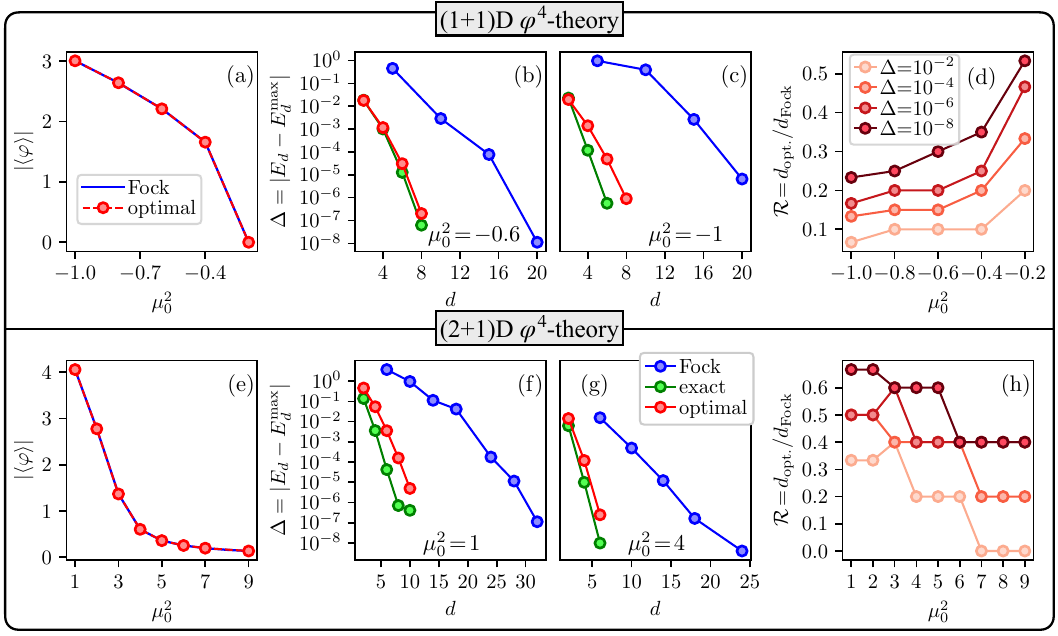}
\caption{\textbf{Local basis optimization for the $\field^{4}$-theory.}
Results for the $(1\!+\!1)\mathrm{D}$ case are shown in the upper row with coupling $\lambda=0.6$ on a  $N=64$-site chain, while the lower row presents the $(2\!+\!1)\mathrm{D}$ case with coupling $\lambda=0.6$ on a $\Lambda\!=\!8\times8$ lattice.  
(a,f) Expectation value of the field operator $\braket{\field}$ as a function of the bare mass $\mu_{0}^2$ obtained using the Fock basis (blue) and the optimal (red), showing good agreement.  
(b–c,e–g) Convergence of the ground-state energy $E_{d}$ as a function of the local Hilbert-space dimension $d$. 
We compare the performance of the Fock (blue), the exact (green), and the optimal (red) bases. 
In the $(1\!+\!1)\mathrm{D}$ case, the optimal basis is derived from an 8-site MF calculation, for bare masses (b) $\mu_{0}^2\!=\!-0.6$ and (c) $\mu_{0}^2=-1$.
In the $(2\!+\!1)\mathrm{D}$ case, the optimal basis is obtained from the mean RDM on a lattice $\Lambda\!=\!4\times 4$, with (e) $\mu_{0}^2\!=\!1$ and (g) $\mu_{0}^2\!=\!4$.  
(d,h) Ratio $\mathcal{R}=d_{\text{opt.}}/d_{\text{Fock}}$ between the minimal local (optimal and Fock basis) dimensions required to achieve a target energy accuracy of $\Delta \in \{10^{-2},10^{-4},10^{-6},10^{-8}\}$ as a function of the coupling $\mu_{0}^{2}$.
}
\label{fig3_phi4}
\end{figure*}

In \cref{fig2_sinegordon}, we study how the ground-state energy $E_{d}$ converges as a function of the local Hilbert space dimension $d$ for different choices of local bases: (i) the standard Fock basis, which does not rely on any prior knowledge of the system;
(ii) an exact basis, constructed a posteriori from the RDM of reference calculation performed in the Fock basis ($\Lambda_{\mathrm{max}}\!=\! 32$-chain, $d_{\mathrm{max}}$, $\chi_{\mathrm{max}}\!=\!100$).
Such an exact basis is used as a theoretical benchmark (it is directly informed by the most accurate wavefunction available) to quantify the most precise ground-state estimate.
Finally, (iii) an \emph{optimal} basis, generated from the RDM of a 3-site MF calculation.
As shown in \cref{fig2_sinegordon}(b–c), the \emph{optimal} basis achieves a convergence rate $\Delta\!\equiv\!\abs{E_{d}\!-\!E_{d}^{\text{max}}}$ comparable to the \emph{exact} one while only requiring the solution of an MF problem, with a much lower computational cost.

More generally, across the whole broken phase, the optimal basis provides highly accurate ground-state energy predictions, with a local dimension $d_{\text{opt.}}$ much smaller than the original Fock basis $d_{\text{Fock}}$.
The corresponding ratio $\mathcal{R}\!=\!d_{\text{opt.}}/d_{\text{Fock}}$ is shown in \cref{fig2_sinegordon}(d) for different target errors $\Delta \!\in\!\{10^{-2},10^{-4},10^{-6},10^{-8}\}$. 
For large $m_{0}^{2}$, the order parameter saturates and correspondingly the ratio $\mathcal{R}$ approaches a small yet finite value.

\subsection{The \texorpdfstring{$\field[][4]$}{phi4} theory}
\label{sec_phi4}
A similar analysis can be performed for the $\field[][4]-$ theory.  
This model has been extensively studied using numerical simulations, including Monte Carlo techniques \cite{Loinaz1998MonteCarloSimulation}, DMRG \cite{Sugihara2004DensityMatrixRenormalization}, and TN algorithms \cite{Milsted2013MatrixProductStates-1, Delcamp2020ComputingRenormalizationGroup, Vanhecke2022EntanglementScaling$ensuremathlambdaensuremathphi_2^4$, Kadoh2019TensorNetworkAnalysis}.  
The discretized lattice Hamiltonian reads
\begin{equation}
    \Ham = \sum_{\vecsite,k} \qty[
        \frac{\mom[\vecsite]^{2}}{2} 
        + \frac{\qty(\field[\vecsite+\latvec[k]]\!-\!\field[\vecsite])^{2}}{2} 
        + \frac{\mu_{0}^2}{2} \field[\vecsite][2]
        + \frac{\lambda}{4!} \field[\vecsite][4]
    ],
    \label{eq_phi4_hamiltonian}
\end{equation} 
The Hamiltonian is invariant under the discrete $\mathbb{Z}_2$ symmetry $\field[\vecsite] \rightarrow -\field[\vecsite]$, and the continuum limit displays an ultraviolet divergence that can be absorbed by redefining the mass parameter as
\begin{math}
    \mu_{R}^2=\mu_{0}^2-\delta \mu^2,
\end{math}
where $\delta \mu^2$ is a counter-term to subtract the divergence.

The $(1\!+\!1)\mathrm{D}$ $\field[][4]$-theory exhibits a second-order quantum phase transition in the $2D$ Ising universality class \cite{Caginalp1980ThermodynamicPropertiesf4,Dash1976UniversalityTwistedFans}, driven by the spontaneous breaking of its $\mathbb{Z}_2$ symmetry. 
This transition separates a \emph{symmetric} phase from a spontaneously \emph{broken} phase and is controlled by the ratio of couplings $\lambda/\mu_R^2$.
The order parameter is given by the expectation value of the field operator $\field$
\begin{equation}
    \braket{\field}= \frac{1}{N} \sum_{\vecsite \in \Lambda} 
    \braket{\field[\vecsite]} 
    =A(\lambda)
    \left( 
    \frac{\lambda}{\mu_{R}^2}
    - \frac{\lambda_c}{\mu^2_{R,c}(\lambda)}
    \right)^{\beta(\lambda)},
    \label{eq: phi4_vev}
\end{equation}
where $A(\lambda)$ is a constant for given coupling $\lambda$. 
In the regime where $\braket{\field}\neq 0$, the model is in the broken phase, and when $\braket{\field}=0$, the model is in the symmetric phase. 
The critical coupling $(\lambda/\mu_{R}^2)_{c}$ and the critical exponents are defined in the continuum limit $N\rightarrow\infty, \lambda\rightarrow 0$.
To illustrate the phase diagram at finite system size, in \cref{fig3_phi4}(a) we show the scaling of the order parameter as a function of the bare mass $\mu^2_{0}$, for fixed coupling $\lambda=0.6$, obtained from MPS simulations on a $N\!=\!64$-site chain with bond dimension $\chi\!=\!100$.

Similar to the Sine-Gordon model, the challenges of a large local Hilbert space become most severe in the \emph{broken} phase, where the expectation value $\braket{\field}$ increases with the power of $\beta(\lambda)$. 
Consequently, the number of Fock levels required in the truncation must increase
accordingly.

This can be seen in \cref{fig3_phi4}(b–c), where we examine the convergence of the ground-state energy as a function of the local dimension. 
We find that the optimal basis, obtained from an 8-site MF calculation, achieves a convergence rate comparable to the exact basis one obtained with an $N\!=\!64$-site chain.
As shown in \cref{fig3_phi4}(d), the deeper in the broken phase, the smaller the ratio $\mathcal{R}\!=\!d_{\text{opt.}}/d_{\text{Fock}}$ and therefore, the larger the advantage of the LBO protocol.

A similar analysis is performed for the $(2+\!1)\mathrm{D}$ $\field^{4}$-theory, with the same lattice Hamiltonian in \cref{eq_phi4_hamiltonian}, $\Lambda$ being a $2\mathrm{D}$ square lattice.
In this case, we investigate the LBO protocol also in the symmetric (non-broken) phase.
In two spatial dimensions, the continuous model exhibits a second-order phase transition in the $3D$ Ising universality class \cite{Henkel1987FiniteSizeScaling, Guida19973DIsingModel, Caginalp1980ThermodynamicPropertiesf4, Dash1976UniversalityTwistedFans}, characterized by the spontaneous symmetry breaking of its $\mathbb{Z}_{2}$ symmetry. 
The phase transition is again detected via the field expectation value defined in~\cref{eq: phi4_vev}, which we plot in absolute value in~\cref{fig3_phi4}(e) as a function of the bare mass $\mu^2_{0}$ at a fixed coupling $\lambda=0.6$. 
The numerical calculations are performed with TTN states on a $\Lambda=8 \times 8$ lattice.
As shown in \cref{fig3_phi4}(f-g), similarly to the ($1\!+\!1$)D case, the optimal basis, obtained from the mean RDM of a TTN simulation on a $\Lambda=4\!\times\!4$ lattice, provides a convergence rate $\Delta$ faster than the Fock basis and comparable to the exact one.
The advantage with respect to the Fock basis, expressed with the ratio $\mathcal{R}\!=\!d_{\text{opt.}}/d_{\text{Fock}}$, is visible in \cref{fig3_phi4}(h) across any value of the bare mass $\mu_{0}^{2}$ at fixed coupling $\lambda\!=\!0.6$.
As expected, the local dimension required in the broken phase ($\mu_{0}^2\!<\!\mu_{0,\text{crit.}}^2\!\sim\!3$) is larger than in the unbroken phase ($\mu_{0}^2\!>\!\mu_{0,\text{crit.}}^2$), but, the most significant improvement from the LBO scheme (\idest{} a lower ratio $\mathcal{R}$) is observed in the non-broken regime. 
For instance, at $\mu_{0}^2=9$, the local dimension $d$ can be reduced by  $\sim\!60\%$ ($\mathcal{R}\!=\!0.4$) for a target accuracy $\Delta = 10^{-8}$, or even by $\sim\!95\%$ ($\mathcal{R}\!=\!0.05$) for $\Delta\!=\!10^{-2}$.

\subsection{Basis augmentation}
\label{sec_basis_augmentation}
\begin{figure}[!t]
    \centering
    \includegraphics[width=0.4\textwidth]{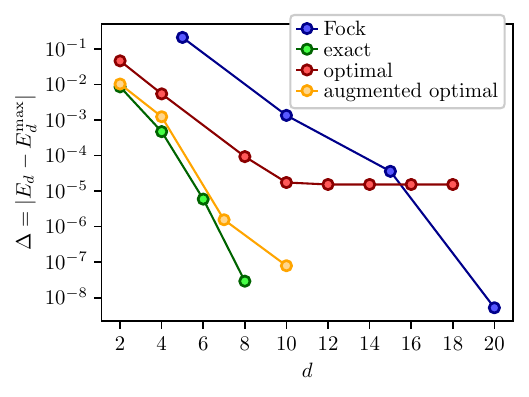}
    \caption{\textbf{Effect of augmentation on the optimal basis.} Convergence of the ground-state energy $E_d$ of the $(1\!+\!1)\mathrm{D}$ $\varphi^4$-model at $\mu_{0}^{2}=-0.6, \lambda=0.6$, with increasing local dimensions in different bases: the Fock basis, the exact basis, the optimal basis, obtained from 2-site MF calculation, and its augmented version. 
    The augmented optimal basis achieves a convergence rate comparable to that of the exact basis, while also reaching the target accuracy of $\Delta\!=\!10^{-8}$. 
    All MPS calculations are performed on a $N\!=\!64$-site chain with bond dimension $\chi\!=\!100$.}
    \label{fig4_augmentation}
\end{figure}
Whenever the local basis obtained from the RDM, computed via MF or TN methods, fails to capture all relevant local modes (typically due to the lack of long-range entanglement in these calculations), the basis augmentation protocol described in \cref{sec_LBO_scheme} can be employed.
In this section, we demonstrate its application in the \emph{broken} phase of the $(1\!+\!1)\mathrm{D}$ $\field^4$-theory, which, as discussed in \cref{sec_phi4}, particularly benefits from the LBO protocol.

The augmentation provides a systematic way to assess the quality of the estimated RDM and improve the local basis. 
By applying local, non-diagonal operators of the theory to the $k$ most significant RDM eigenstates, the local basis is extended to include modes typically missed in MF or finite-size TN calculations.
This procedure enriches the basis with both lower- and higher-energy local modes surrounding the initially captured ones (see \cref{fig1_scheme}(d)), replicating the effect of increasing the Fock space.

The application of the method is illustrated in \cref{fig4_augmentation}, where we show the convergence of the ground-state energy with increasing local Hilbert space dimension for four different bases: the Fock basis, the exact basis, the optimal basis (obtained from the a 2-site cluster-MF calculation), and the augmented basis constructed by extending the most significant eigenvectors of the optimal basis. 
The augmentation is performed by iteratively applying an operator from the set of local operators appearing in the Hamiltonian $\{\mom[i]^2, \field[i]^2, \field[i]^4\}$. 
After the application of each operator, we assess, through~\cref{eq_tolerance_augmentation}, whether the resulting states are sufficiently orthogonal (up to a chosen tolerance $r\!=\!10^{-8}$) to the previous optimal basis. 
If this condition is satisfied, the optimal basis is extended and the Gram-Schmidt algorithm is applied.
We observe that, close to the critical point ($\mu_{0}^{2}=-0.6, \lambda=0.6$), the convergence in energy with the optimal basis saturates, even when including eigenstates with small eigenvalues (below $\!<\!10^{-16}$). 
This indicates that the 2-site MF estimate of the RDM fails to approximate the exact RDM.
In contrast, the proposed augmentation method leads to continued improvement in the ground-state energy, reaching the target accuracy $\Delta\!=\!10^{-8}$. 
The convergence behavior is comparable to that achieved using the exact basis.
All results are obtained from MPS simulations on an $N\!=\!64$-site chain at bond dimension $\chi\!=\!100$.

\begin{figure*}[ht]
    \centering
    \includegraphics[width=1\textwidth]{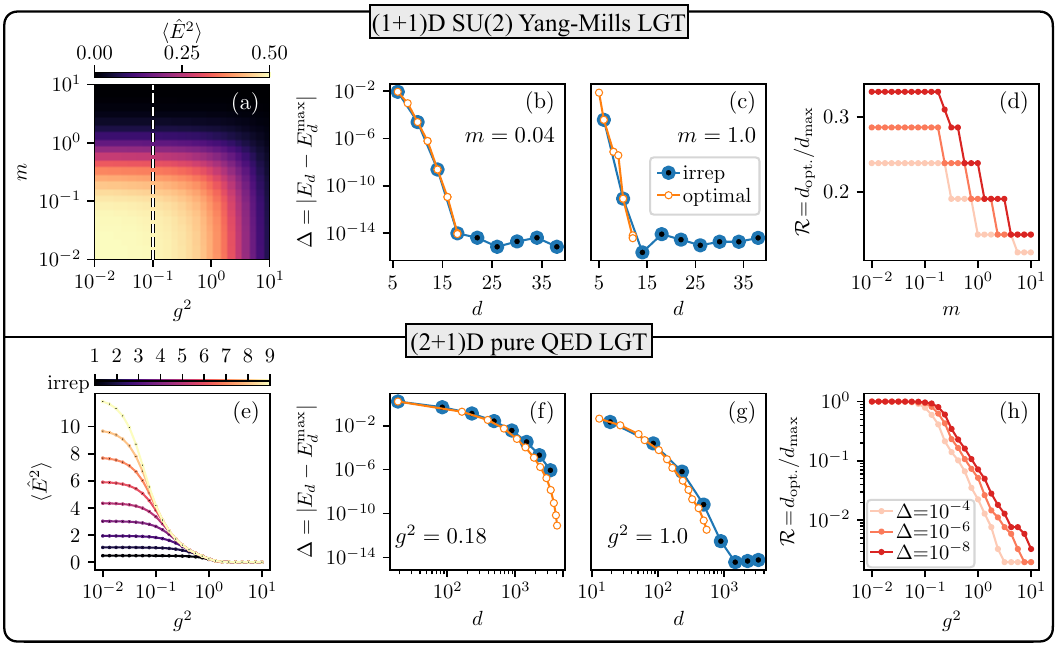}
\caption{\textbf{Local basis optimization in Lattice Gauge Theories.}
Results obtained from ED simulations of (first row) $(1\!+\!1)\mathrm{D}$ SU(2) Yang-Mills LGT with dynamical matter on a $N\!=\!10$-site chain in PBC, and (second row) $(2\!+\!1)\mathrm{D}$ pure QED LGT on a $\Lambda\!=\!2\!\times\!2$ lattice in PBC.
(a,e) Expectation value of the Casimir operator $\braket{\hat{E}^{2}}$ across the phase diagram: in the plane spanned by the bare mass $m$ and the coupling $g^2$ for SU(2), and as a function of $g^2$ for QED.
(b–c,f–g) Convergence of the ground-state energy $E_{d}$ as a function of the local Hilbert-space dimension $d$. 
We compare the performance of the dressed site formulation built using the electric (irrep) basis with that of the optimal one obtained from the single-site RDM.
(d,h) Scaling of the ratio $\mathcal{R}\!=\!d_{\text{opt.}}/d_{\max}$ as a function of the coupling $m$ (for SU(2)) and $g^{2}$ (for QED) at a fixed target energy accuracy $\Delta\in \{10^{-4},10^{-6},10^{-8}\}$, where the reference local dimension $d_{\max}$ obtained from the maximal irrep basis ($\spin=5$ for SU(2) and $\spin=9$ for QED) in the dressed-site formulation.}
\label{fig5_LGT}
\end{figure*}
\section{Lattice Gauge Theories}
\label{sec_LGTs}

We test the LBO protocol on two LGTs: the $(1\!+\!1)\mathrm{D}$ SU(2) Yang-Mills LGT with flavorless dynamical matter \cite{Calajo2024DigitalQuantumSimulation-1, Cataldi2025DisorderFreeLocalizationFragmentation, Calajo2025QuantumManybodyScarring} and the $(2\!+\!1)\mathrm{D}$ pure quantum electrodynamics (QED) \cite{Felser2020TwoDimensionalQuantumLinkLattice-1, Magnifico2021LatticeQuantumElectrodynamics-1, Magnifico2025TensorNetworksLattice-1} (without dynamical matter).
In both cases, we adopt the \emph{Kogut-Susskind} formulation \cite{Kogut1975HamiltonianFormulationWilsons-1}, while dynamical matter is described using \emph{staggered fermions} \cite{Wilson1974ConfinementQuarks-1}. 
The gauge degrees of freedom, which in principle belong to an infinite local Hilbert space, are decomposed in the eigenbasis of the Casimir (electric or irrep basis) and truncated in terms of a fixed Casimir eigenvalue.
To locally enforce Gauss's law (in both U(1) and SU(2) cases), we employ the dressed-site scheme~\cite{Silvi2014LatticeGaugeTensor-1, Magnifico2025TensorNetworksLattice-1, Cataldi2024Simulating2+1DSU2-1}, in which each fermionic matter site is fused with its neighboring bosonic gauge half links, resulting from splitting each local gauge field, into a (bosonized) basis consisting solely of gauge-invariant combinations of the fields.
For details, see~\cref{app_dressed_site}. 
To measure convergence with respect to the local dimension, we measure the Casimir operator.

\subsection{(1+1)D SU(2) Yang-Mills LGT}
\label{sec: SU2}
The $(1\!+\!1)\mathrm{D}$ SU(2) Yang-Mills LGT with flavorless dynamical matter reads (in units of lattice spacing) \cite{Calajo2024DigitalQuantumSimulation-1, Calajo2025QuantumManybodyScarring, Cataldi2025DisorderFreeLocalizationFragmentation}:
\begin{equation}
    \label{eq_H_su2}
    \begin{split}
    H_{\rm{SU(2)}}^{\rm{1D}}&=\frac{1}{2}\sum_{\vecsite}\sum_{\alpha, \beta}
    \qty[i\hpsi^{\dagger}_{\vecsite,\alpha}\hat{U}^{\alpha\beta}_{\vecsite,\vecsite+\latvec[x]}\hpsi_{\vecsite+\latvec[x],\beta}+\hc] \\
    &+ m\sum_{\vecsite,\alpha} (-1)^{\vecsite} \hpsi^{\dagger}_{\vecsite,\alpha}\hpsi_{\vecsite,\alpha}
    + \frac{g^2}{2}\sum_{\vecsite}\hat{E}^2_{\vecsite,\vecsite+\latvec[x]}
    \,. 
    \end{split}
\end{equation}
Here, fermionic matter degrees of freedom $\matt{\vecsite}$ represent are located on the lattice sites $\vecsite$ and obey the anti-commutation relation
\begin{math}
  \{\matt{\vecsite}, \matt*{\vecsite^{\prime}}[\beta]\} = \delta_{\alpha\beta}\delta_{\vecsite\vecsite^{\prime}},
\end{math}
where $\alpha$ and $\beta$ are color indices.
The gauge degrees of freedom, residing on the lattice links, are described by the parallel transporter $\para{\vecsite}$, which transforms under the fundamental representation, and the Casimir operator $\Op{E}^2$, which corresponds to the electric energy (no magnetic term is present in $(1\!+\!1)\mathrm{D}$). 
The first two terms of \cref{eq_H_su2} represent the matter-gauge interaction (hopping) and the local staggered mass term. 

Using ED methods, we simulate a $N\!=\!10$-site chain with periodic boundary conditions (PBC) and 10 incremental truncations of the $SU(2)$ gauge field, from $\jmax\!=\!\jonehalf$ (with a local dressed-site basis of $d\!=\!6$ sites~\cite{Calajo2024DigitalQuantumSimulation-1}) to $\jmax\!=\!5$ (with local dimension $d\!=\!42$), the latter adopted for the reference results.
We compute the ground-state model phase diagram for a grid of couplings $m, g^{2}\in\qty[10^{-2},10^{1}]$ shown in \cref{fig5_LGT}(a).

In the absence of the magnetic term, any fluctuation of the gauge field irrep is obtained via the hopping term, which in turn is dominant only for small $m$.
Therefore, the model undergoes a transition in $m$ that can be detected through local observables, such as the Casimir (but also via particle densities).
The small-$m$ phase and the transition depend on the adopted gauge-irrep truncation $\jmax$ and hence on the local dimension $d$. 

Fixing $g^{2}=0.1$ in \cref{fig5_LGT}(a) and adopting the LBO scheme proposed in \cref{sec_LBO_scheme}, we compare the performances of the standard electric basis with the corresponding one obtained from the single-site RDM truncated with different precision errors $\varepsilon$. 
As shown in \cref{fig5_LGT}(b,c), the two bases provide the same results, and incremental irreps correspond to incremental truncation errors in the single-site RDM.
Essentially, the electric basis proves to be the optimal local basis for such $(1\!+\!1)\mathrm{D}$ non-Abelian LGT at any regime of the mass and gauge coupling.
Indeed, at the ground-state level, the irreps fusing in the Gauss Law are not mixed and contribute to the energy in ascending order with the corresponding Casimir, which is reflected in the single-site RDM.
Moreover, as shown in \cref{fig5_LGT}(d), across all the simulated couplings, $\jmax\!=\!3/2$, corresponding to $d\!=\!14$, guarantees convergence in observables with error $\varepsilon\leq10^{-8}$, extending what was already known in the Abelian counterpart \cite{Buyens2017FiniterepresentationApproximationLattice-1} and confirming the validity of interesting $(1\!+\!1)\mathrm{D}$ non-Abelian dynamical features recently investigated \cite{Calajo2024DigitalQuantumSimulation-1, Calajo2025QuantumManybodyScarring, Cataldi2025DisorderFreeLocalizationFragmentation}.

\subsection{(2+1)D pure QED LGT}
\label{sec: 2d_pure_qed}
To further investigate the local basis optimization scheme in LGTs, we consider QED in two spatial dimensions, the minimal example where magnetic effects contribute and can affect the mixing of gauge link irreps.
For simplicity, we focus on the pure theory without dynamical matter. 
The corresponding $(2\!+\!1)\mathrm{D}$ pure QED Hamiltonian reads:
\begin{equation}
\label{eq_H_QED}
  \hat{H}_{\rm{QED}}^{\rm{2D}}=\frac{g^2}{2}\sum_{\vecsite,k}
  \hat{E}^{2}_{\vecsite,\latvec[k]}-\frac{1}{2g^2} \sum_{\square} \mathrm{Tr} (\hat{U}_{\square}+\hat{U}^{\dagger}_{\square}).
\end{equation}
Here, in addition to the electric term, expressed in terms of the Casimir operator $\hat{E}^{2}$, we have a magnetic term expressed by the square plaquette operator $\hat{U}_{\square}$:
\begin{equation}
  \hat{U}_{\square} = \hat{U}_{\vecsite,+\vb*{\mu}_{x}}\hat{U}_{\vecsite+\vb*{\mu}_{x},+\vb*{\mu}_{y}}\hat{U}_{\vecsite+\vb*{\mu}_{y}+\vb*{\mu}_{x},-\vb*{\mu}_{x}}\hat{U}_{\vecsite,+\vb*{\mu}_{y}},
\end{equation}
where $\vb*{\mu}_{x}$ and $\vb*{\mu}_{y}$ denote the lattice unit vectors spanning the plaquette’s plane.

Similarly to the SU(2) case, we adopt the dressed site formulation with gauge link truncation based on incremental energy cutoffs in the Casimir and consider sequential spin-discretizations of the gauge fields, ranging from $\spin=1$ (with a local dressed-site basis of $d\!=\!19$ sites~\cite{Felser2020TwoDimensionalQuantumLinkLattice-1}) to $\spin=9$ (with $d\!=\!4579$).
With ED methods \cite{Cataldi2024EdlgtExactDiagonalizationLattice-2}, we simulate a lattice $\Lambda\!=\!3\!\times\!2$ in PBC on a grid of couplings $g^{2}\in \qty[10^{-2}, 10^{1}]$.
In this range of the couplings, the model in \cref{eq_H_QED} undergoes a transition between a large-$g$ phase, dominated by the electric term, where gauge field excitations are energetically expensive and then suppressed, and a small-$g$ phase, dominated by the magnetic term, which favors and makes gauge field excitations cheaper.
As shown in the \cref{fig5_LGT}(e), the shape of such a transition is strongly affected by the gauge fields truncation, while the electric phase is well described already at small truncations \cite{Buyens2017FiniterepresentationApproximationLattice-1}.

As shown in~\cref{fig5_LGT}(f-g), similarly to the $(1\!+\!1)\mathrm{D}$ SU(2) case, the exact local basis coincides everywhere with the corresponding electric one, signalling that, at the single-site level, the electric basis confirms as a solid strategy.
At the same time, in $(2\!+\!1)\mathrm{D}$, the small-$g$ phase (\idest{} where the continuum limit should be located~\cite{Magnifico2025TensorNetworksLattice-1}) is composed of highly entangled single-sites, preventing significant truncations of the local basis (see \cref{fig5_LGT}(h)). 

This motivated the need for an alternative \emph{magnetic} basis formulation, where the magnetic term (dominating in the small-$g$ phase) becomes local in a dual lattice \cite{Kaplan2020GausssLawDuality-1} and reduces the source of entanglement. 
Promising proposals have been addressed in the Abelian \cite{Haase2021ResourceEfficientApproach-1, Paulson2021Simulating2DEffects-1} and non-Abelian \cite{Fontana2024EfficientFiniteresourceFormulation} scenarios, while remaining so far restricted to small system size simulations and carrying on long-range Hamiltonian terms.

\begin{figure}[!t]
    \centering
    \includegraphics[width=1\columnwidth]{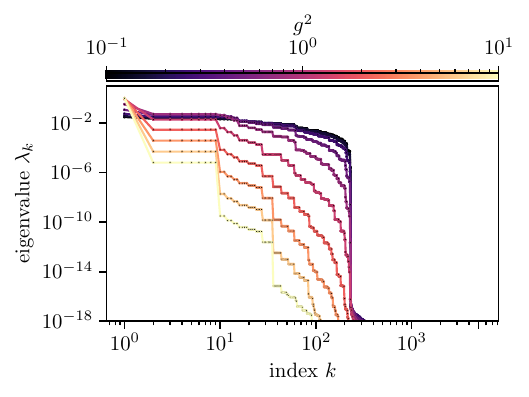}
    \caption{\textbf{QED single-plaquette RDM eigenvalues.} Scaling of ground-state plaquette RDM eigenvalues of the $(2\!+\!1)\mathrm{D}$ QED pure LGT in \cref{eq_H_QED} as a function of the gauge coupling $g^{2}$. 
    Results are obtained for ED methods on a $3\!\times\!2$ lattice in PBC with $\spin_{\max}=3$ (\idest{}, local dimension $d=231$).}
    \label{fig6_LGTsvd}
\end{figure}

In this perspective, the LBO scheme proposed in this work can provide an alternative solution. 
Indeed, in the $(2\!+\!1)\mathrm{D}$ pure QED, already at $\jmax\!=\!3$ (with local dimension $d\!=\!231$), the eigenvalues of single-plaquette RDM (whose gauge invariant Hilbert space has $5299$ states) exhibit a significant decay across all the $g$ values (see \cref{fig6_LGTsvd}), which is not visible in the single-site RDM. 
This suggests that, starting from a single-site electric basis, we can project the Hamiltonian onto an \emph{optimal plaquette basis} and obtain a more efficient representation of the local degrees of freedom with a drastic reduction of computational cost.
At small-$g$ values, we expect at least an order of magnitude improvement (from $5299$ to $\sim\!250$). 
At intermediate $g$-values, such an improvement reaches two orders of magnitude.
A systematic exploration of the resulting computational gains will be addressed in future work.

\section{Improved resource estimate of classical and quantum computing}
\label{sec_quantum_simulation}
In this section, we discuss the practical implications of our LBO scheme for classical simulations across different TN architectures, as well as for digital quantum computing.

\subsection{Resource estimate for classical simulations}
The LBO procedure previously developed applies broadly across ED, TN, MF, and cluster-MF simulations, being entirely problem-independent. 
Although the computational complexity of a given TN ansatz may vary, reducing the local dimension $d$ enhances efficiency in all cases, resulting in significant runtime improvements.

For instance, for the MPS-based DMRG algorithm, among the most accurate methods for one- and quasi-one-dimensional QMB systems, the computational cost for ground-state calculations scales as $\mathcal{O}\left(N d\chi^3\!+\!N d^2 \chi^2\right)$ or $\mathcal{O}\left(N d^3 \chi^3\right)$, depending on whether single- or 2-site updates are used during the optimization sweeps ~\cite{Schollwock2011DensitymatrixRenormalizationGroup-1, Chan2016MatrixProductOperators-1}. 

In higher dimensions, DMRG algorithms on PEPS, the natural generalization of MPS, scale as $\mathcal{O}\left(N d^2 \chi^8\right)$ \cite{Eisert2013EntanglementTensorNetwork-1, Vanderstraeten2022VariationalMethodsContracting-1}.
This limits simulations to relatively small bond dimensions $\chi\!\approx\!10$. 
Here, an \emph{optimal} local basis is even more critical, particularly for bosonic systems, as it can render simulations feasible that would otherwise be computationally prohibitive.
As for binary TTN architecture, ground-state searches scales with $\mathcal{O}\left(N d^2\chi^2 + N \chi^4\right)$, regardless of the 
spatial dimension~\cite{Silvi2019TensorNetworksAnthology-1}. 

An optimal basis is also crucial in TN algorithms for time evolution. 
In TEBD (time evolving block decimation)\cite{Vidal2003EfficientClassicalSimulation-1, Vidal2004EfficientSimulationOnedimensional-1}, the local dimension scales at least quadratically in any matrix product operator operation on MPS~\cite{Paeckel2019TimeevolutionMethodsMatrixproduct-1}. 
There, the LBO acts as a projection on each physical leg of the MPS, mapping it to a smaller basis (see also \cite{Brockt2015MatrixproductstateMethodDynamical-1}). 
In TDVP (time-dependent variational principle)~\cite{Haegeman2011TimedependentVariationalPrinciple-1}, the LBO scheme applies directly as for DMRG \cite{Schroder2016SimulatingOpenQuantum-1}.

More generally, while the scaling with $d$ can be either leading or sub-leading with respect to the bond dimension $\chi$, reducing $d$ consistently leads to substantial performance improvements, particularly in systems where $d$ is very large. 
This is especially relevant for LGTs, where local dimensions as large as $d \geq 10^{6}$ are possible depending on the truncation scheme \cite{Magnifico2025TensorNetworksLattice-1}.
Basis optimization thus emerges as a universally applicable strategy for reducing computational costs across all TN ansätze, enabling simulations of large-scale QMB systems.

\subsection{Resource estimate for quantum computing}
In this section, we estimate the impact of our LBO protocol on digital quantum simulations of field theories. 
A typical digital quantum simulation involves three main tasks: \emph{(i)} preparing an initial state, \emph{(ii)} simulating the time evolution, \emph{(iii)} and measuring final observables. 
Our method affects all these steps by providing a scheme to find an optimal realisation for the field configurations. 
For simplicity, we focus on the representation in terms of qubits, but there exist proposals to simulate field theories with qudit systems~\cite{Kurkcuoglu2021QuantumSimulationPhi^4}, which would be similarly affected.

A state of the lattice field theory is represented by assigning to each lattice site a register of qubits, which truncates the unbounded field variable to a maximum value $\tilde{\varphi}_\text{max}$ and discretizes it with resolution $\Delta_{\tilde{\varphi}}$. 
This low-energy subspace at lattice site $\vecsite$ can then be represented as the eigenspace of the discrete field operator $\tilde{\Phi}$ \cite{Macridin2022BosonicFieldDigitization, Li2023SimulatingScalarField}
\begin{equation}
    \tilde{\Phi} \ket{\tilde{\varphi}_{\vecsite}}=\tilde{\varphi}_{\vecsite}\ket{\tilde{\varphi_{\vecsite}}}, \qquad
    \tilde{\varphi}_{\vecsite} = \vecsite\, \Delta_{\tilde{\varphi}}
\end{equation}
where $\vecsite\!=\!-\!\tfrac{N_{\tilde{\varphi}-1}}{2}, \tfrac{N_{\tilde{\varphi}-1}}{2}\!+\!1,\dots, \tfrac{N_{\tilde{\varphi}-1}}{2}$ 
and  
\begin{math}
    \Delta_{\tilde{\varphi}}
    \!=\!\sqrt{\frac{2\pi}{N_{\tilde{\varphi}}\mu_{0}^2}}.
\end{math}
To prepare a Gaussian state $\ket{\psi}$ of the \emph{free} $\field^{4}$-model ($\lambda\!=\!0$), a field cut-off
$\tilde{\varphi}_{\text{max}}=\sqrt{\frac{|\lambda|E}{\mu_{0}^{2}\epsilon}}$ is sufficient to achieve fidelity of $1\!-\!\epsilon$
\cite{Jordan2012QuantumAlgorithmsQuantum}. 
The discrete conjugated momentum field operator $\tilde{\Pi}$ is then obtained, through the discrete Fourier transform $\tilde{\mathcal{F}}$ as 
\begin{math}
    \tilde{\Pi}= \tilde{\mathcal{F}} \tilde{\Phi} \tilde{\mathcal{F}}^{-1},
\end{math}
defined as 
\begin{align}
    \mathcal{\tilde{F}}= \frac{1}{\sqrt{N_{\tilde{\varphi}}}} 
    \sum_{\vecsite,\vecsite^{\prime}=-M}^{M}
    e^{\frac{2 \pi i}{N_{\tilde{\varphi}}}} \ket{\tilde{\varphi}_{\vecsite}}\bra{\tilde{\varphi}_{\vecsite^{\prime}}},
\end{align}
where $M\!=\!\frac{N_{\tilde{\varphi}}-1}{2}$ is the range of summation.
The momentum eigenstates $\ket{\tilde{p}_{\vecsite}}$ of the discrete momentum operator $\tilde{\Pi}$ are then constructed by applying the Fourier transform to the eigenstates of the discrete field operator:
\begin{math}
    \ket{\tilde{p}_{\vecsite}}\!=\!\mathcal{\tilde{F}}\ket{\tilde{\varphi}_{\vecsite}}.
\end{math}
The states generators of the field operators $\field, \mom$ of the free $\field^{4}$-theory (restricted to the lowest 
$N_{b}$ bosonic excitation) are then identical to the $N_{b}$ bosonic excitation of the restricted operators $\tilde{\Phi}, \tilde{\Pi}$ up to an error $\epsilon$, which decreases exponentially with $N_{\tilde{\varphi}}$~\cite{Macridin2022BosonicFieldDigitization}. 
Hence, the larger the truncation $\tilde{\varphi}_{\text{max}}$ and the finer the field discretization, the better the agreement between the bosonic excitation generated by the discrete fields $\tilde{\Phi}$ and $\tilde{\Pi}$ and that of the free $\field^4$-theory.

\begin{figure}[!t]
    \centering
    \includegraphics[width=0.4\textwidth]{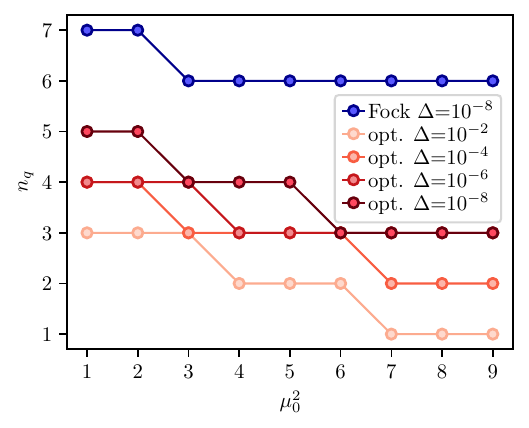}
    \caption{\textbf{Qubit resource estimate.}
    Number of qubits required $n_{q}$ as a function of the bare mass $\mu_{0}^{2}$, obtained by applying the ceiling function to~\cref{eq: qubit per side}, 
    to represent the field operator $\field[\vecsite]$ for different truncations $\Delta \in \{10^{-2},10^{-4},10^{-6},10^{-8}\}$. 
    The results correspond to the $\field^4$-theory in $(2\!+\!1)\mathrm{D}$ on a lattice $\Lambda = 8 \times 8$ with fixed coupling $\lambda=0.6$.
    }
    \label{fig_phi4_number_qubits}
\end{figure}

To store the local field configuration $\ket{\tilde{\varphi}_{\vecsite}}$, a qubit register of size 
\cite{Jordan2012QuantumAlgorithmsQuantum}
\begin{equation}
    n_{q} = \mathcal{O}\left(\log_{2}\left(\tilde{\varphi}_{\text{max}}/
    \Delta_{\tilde{\varphi}}\right)\right)
    \label{eq: qubit per side}
\end{equation}
is required. In practice, instead of expanding the field variables in the eigenvectors of the discretized field operators $\tilde{\Phi},\tilde{\Pi}$, they are expanded in the bosonic occupation number basis, as described in \cref{sec_scalar_field_theories}, i.e., the eigenstates of the discretized local harmonic oscillator and the local Hilbert space is truncated by the boson occupation number.
It was shown in \cite{Macridin2022BosonicFieldDigitization} that the low-energy subspace, up to a boson occupation number $N_{b}$, provides an efficient representation with an accuracy $\mathcal{\epsilon}$ of the low-energy subspace of the discretized field operators for $N_{\tilde{\varphi}}> N_{b}.$ 
The accuracy $\mathcal{\epsilon}$ improves exponentially with increasing cutoffs $N_{\tilde{\varphi}}$.
In~\cref{fig_phi4_number_qubits}, we show how the size of the qubit register $n_{q}$ necessary to represent a field configuration $\tilde{\varphi}_{j}$ varies, and we plot it throughout the phase diagram as a function of the bare mass, at a fixed coupling $\lambda = 0.6$, and for different target accuracy $\Delta\!\in\!\{10^{-2},10^{-4},10^{-6},10^{-8}\}$. 
The calculations were performed for the $\field^4-$ model in $(2\!+\!1)\mathrm{D}$, on a $\Lambda= 8 \times 8$ lattice. 
For both the Fock and the optimal basis, the number of qubits required to faithfully represent the field increases in the broken phase, for $\mu_{0}^2\!<\!3$. 
However, the most significant reduction in the number of qubits per lattice site is achieved in the unbroken phase, for $\mu_{0}^2\!>\!3$. 
For instance, at $\mu_{0}^2\!=\!8$, the optimal basis needs a single qubit (compared to six needed in the Fock basis) to achieve an accuracy of $\Delta\!=\!10^{-2}$.
We emphasize that the qubit reduction shown in~\cref{fig_phi4_number_qubits} refers to a single lattice site, while the total resource savings scale linearly with the number of lattice sites. 

An alternative approach to quantum simulation is the variational method, exemplified by the quantum approximate optimization algorithm (QAOA)~\cite{Farhi2014QuantumApproximateOptimization} and the variational quantum eigensolver (VQE)~\cite{Peruzzo2014VariationalEigenvalueSolver}. 
For example, VQE approximates the Hamiltonian ground state by preparing a parametrized quantum state with a variational quantum circuit and iteratively optimizing its parameters via classical feedback to minimize the measured energy expectation value. 
Representing a local bosonic degree of freedom with $d\!>\!2$ requires multiple qubits, which induces intrinsically long-range interactions in the Hamiltonian and leads to increased circuit depth. Alternatively, it can be encoded in qudits, but engineering as many transition levels as $d$ is quite demanding already for $d\sim 10$ \cite{Ringbauer2022UniversalQuditQuantum-1}. 
In both scenarios, the proposed LBO protocol can improve the efficiency and accuracy of the VQE procedure.

\section{Conclusions}
\label{sec_conclusion}
We proposed a scheme for constructing an optimal basis that enables an effective description of the target theory. 
The method relies on estimating the single-site RDM, obtained from ED, TN, MF, and cluster MF calculations on small translational-invariant systems.

This approach has been first applied to two paradigmatic scalar field theories: the sine-Gordon model in $(1\!+\!1)\mathrm{D}$ and the $\field^4$-theory in $(1\!+\!1)\mathrm{D}$ and $(2\!+\!1)\mathrm{D}$. 
The benefits of basis optimization are particularly pronounced in symmetry-broken phases, where conventional expansions in the Fock basis typically demand large local Hilbert spaces and thus substantial computational resources. 
The method proves especially advantageous in two spatial dimensions, opening the door to applying TN techniques to bosonic field theories in higher dimensions, where standard approaches are limited by computational cost.

We further extended the method to Abelian and non-Abelian LTGs in $(2\!+\!1)\mathrm{D}$ and $(1\!+\!1)\mathrm{D}$ scenarios, confirming that the electric basis built from incremental spin-irreps of the gauge fields provides an optimal single-site description, particularly in one dimension. 
In $(2\!+\!1)\mathrm{D}$, we investigated the decay of the single-plaquette RDM eigenvalues across different values of the coupling $g^2$. 
Our results show that the optimal dimension of the single-plaquette Hilbert space is significantly smaller than the one required in the original dressed-site formulation of the LGT. 
These findings indicate that expressing the Kogut–Susskind Hamiltonian in a plaquette basis provides a more compact and efficient representation of the local degrees of freedom, especially in the small-$g^2$ regime.

Finally, we explored the advantages of the LBO protocol for digital quantum simulation of the $(2\!+\!1)\mathrm{D}$ $\field^4$-theory. 
We demonstrated that the optimal basis significantly reduces the number of qubits required to encode local field variables throughout the phase diagram.
This reduction is particularly valuable in the current NISQ era \cite{Preskill2018QuantumComputingNISQ-1}, where quantum resources are limited, and provides the most efficient encoding of local degrees of freedom on a quantum device.

\begin{acknowledgments}
The authors acknowledge financial support from the following institutions:
The European Union (EU)
via the Horizon 2020 research and innovation program (Quantum Flagship) projects PASQuanS2 and Euryqa,
via the NextGenerationEU project CN00000013 - Italian Research Center on HPC, Big Data and Quantum Computing (ICSC),
the QuantERA projects T-NiSQ and QuantHEP, and via the Horizon 2020 Research and Innovation Programme under the Marie Skłodowska-Curie Grant Agreement No. 101034267.;
The Italian Ministry of University and Research (MUR)
via the Departments of Excellence grants 2023-2027 projects Quantum Frontiers and Quantum Sensing and Modelling for One-Health (QuaSiModO),
and via PRIN2022 project TANQU;
the PNRR extended partnership National Quantum Science and Technology Institute (NQSTI) via Bandi a Cascata project OPTIMISTIQ;
the Italian Istituto Nazionale di Fisica Nucleare (INFN)
via specific initiatives IS-QUANTUM and IS-NPQCD;
The German Federal Ministry of Education and Research (BMBF)
via project QRydDemo;
the Max Planck Society; the Deutsche Forschungsgemeinschaft (DFG, German Research Foundation) under Germany’s Excellence Strategy – EXC-2111 – 390814868; the European Research Council (ERC) under the European Union’s Horizon Europe research and innovation program (Grant Agreement No.~101165667)—ERC Starting Grant QuSiGauge.;
The World-Class Research Infrastructure $-$ Quantum Computing and Simulation Center (QCSC) of Padova University.
We acknowledge computational resources by the Cloud Veneto, CINECA, the BwUniCluster, and the University of Padova Strategic Research Infrastructure Grant 2017: “CAPRI: Calcolo ad Alte Prestazioni per la Ricerca e l’Innovazione”. 
Views and opinions expressed are, however, those of the author(s) only and do not necessarily reflect those of the European Union or the European Research Council Executive Agency. 
Neither the European Union nor the granting authority can be held responsible for them. 
This work is part of the Quantum Computing for High-Energy Physics (QC4HEP) working group.  
\end{acknowledgments}

\bibliography{bibliography}

\appendix
\section{Models}
\label{app_scalarmodels}
In this section, we provide additional information about the sine-Gordon (sG) and $\field^{4}-$ theories in the continuum; specifically, we give details about their respective renormalization procedure, necessary to extract the properties of the continuum theories from the lattice Hamiltonians in~\cref{eq_sinegordon_hamiltonian,eq_phi4_hamiltonian}.

\subsection{The sine-Gordon theory}
\label{sec_appendix-sG}
The energy density of the continuum quantum sG model in $(1\!+\!1)\mathrm{D}$ is defined as follows \cite{Lukyanov1997ExactExpectationValues, Roy2021QuantumSineGordonModel, Pallua2001UVIRAnalyses}:
\begin{equation}
    \mathcal{H}_{\text{sG}} = \frac{1}{2} \Op{\pi}^2 +
    \frac{1}{2} \partial^{x} \Op{\varphi}\partial_{x} \Op{\varphi}
    - \frac{m_0^2}{\beta^2} \cos(\beta \Op{\varphi}),
\end{equation}
where $m_{0}^2$ is the mass parameter and $\beta^2$ is a dimensionless coupling constant.
The sG model can be interpreted as a deformation of a free conformal field theory (CFT), where the mass term is treated as a perturbation \cite{KLASSEN1993SINEGORDONVSMASSIVE}. 
When evaluated relative to the CFT vacuum, the energy density remains finite and is given by \cite{ZAMOLODCHIKOV1995MASSSCALESINE}
\begin{equation}
    E_0(M) = \frac{\braket{\Ham}_{\text{sG}} - \braket{\Ham}_{\text{CFT}}}{V}
    = -\frac{M^2}{4} \tan\left( \frac{\pi \xi}{2} \right),
\end{equation}
where $M$ is the renormalized soliton mass, and $\xi$ is a parameter depending on the coupling $\beta^2$, defined as $\xi\!=\!\beta^2 /(8\pi\!-\!\beta^2)$. 
The formula for the soliton mass $M$ is 
\cite{Lukyanov1997ExactExpectationValues, Calliari2024QuantumSimulatingContinuum}
\begin{equation}
    M=b\tilde{\Lambda}\frac{2\Gamma(\frac{\xi}{2})}{\sqrt{\pi} \Gamma(\frac{1+\xi}{2})}
    \left(
    \frac{m_{0}^2(1+\xi)\Gamma(\frac{1}{1+\xi}) }{16\xi\Gamma(\frac{\xi}{1+\xi})(b\tilde{\Lambda})^{2}} \right)^{\frac{1+\xi}{2}}.
    \label{eq:sG-soliton_mass}
\end{equation}
Here $\tilde{\Lambda}=\pi/a$ is the UV-cut-off and $b=e^{C}/2$, a parameter depending on the renormalization scheme. 

An alternative way to study the quantum sG model is to use the fact that there exists a spin chain lattice regularization in the form of the XXZ spin chain with periodic boundary conditions in a transverse magnetic field \cite{KLASSEN1993SINEGORDONVSMASSIVE}.

\subsection{\texorpdfstring{$\field[][4]-$}{phi4} theory}
\label{sec_appendix-phi4}
The classical Hamiltonian density of the $\varphi^4-$ model in $(1\!+\!1)\mathrm{D}$ is given by
\begin{align}
        \mathcal{H}_{\varphi^4}
        =&\frac{1}{2} \pi^{2}
        +\frac{1}{2}\partial^{x}\varphi \partial_{x}\varphi 
        +\mu_{0}^2\varphi^{2}
        +\frac{\lambda}{4!}\varphi^{4}.
\end{align}
Here, $\pi$ is the canonical momentum conjugate to the field $\varphi$, defined by
\begin{math}
        \pi=\partial_{t}\varphi.
\end{math}
Both fields $\varphi$ and $\pi$ are dimensionless, while $m$ and $\lambda$ have units of energy. 
The theory is symmetric under the $\mathbb{Z}_2$ transformation: $\varphi \rightarrow -\varphi$.
For $\mu_{0}^2 > 0$, the potential 
\begin{math}
     U(\varphi)=\frac{1}{2}\mu_{0}^2 \varphi^{2}+\frac{\lambda}{4!}\varphi^{4}
\end{math}
has a unique minimum at $\varphi = 0$, preserving $\mathbb{Z}_2$ symmetry. 
For $\mu_{0}^2 < 0$, we can rewrite the potential by adding a constant shift $\Delta U=\frac{3}{2}\frac{|\mu_{0}^2|^{2}}{\lambda}$ as
\begin{align}
    U(\varphi)+\Delta U =\frac{\lambda}{4!}\left(\varphi^{2} -\frac{6 |\mu_{0}^2|}{\lambda}
    \right)^{2},
\end{align}
which has degenerate minima at $\varphi=\pm\sqrt{6|\mu_{0}^2|/\lambda}$. The $\mathbb{Z}_2$ symmetry therefore interchanges the two vacua, resulting in symmetry breaking. 
More importantly, this classical calculation estimates how the local dimension scales as one proceeds further into the broken phase.

\begin{figure}[ht]
    \centering
    \includegraphics[width=0.75\columnwidth]{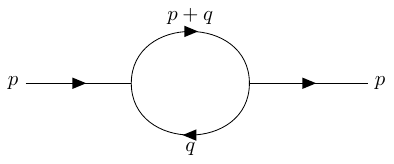}
    \caption{Divergent diagram of the $\field[][4]$ theory in $(1\!+\!1)\mathrm{D}$.}
    \label{fig: Feynman-diagramm}
\end{figure}

After quantization of the $\varphi^4$-theory, as described in \cref{sec_phi4}, the quantum field theory has to be renormalized; this can be done by redefining the mass parameter. 
The counter-term $\delta \mu_{0}^2$ is defined as \cite{ Loinaz1998MonteCarloSimulation,Sugihara2004DensityMatrixRenormalization}
\begin{align}
    \delta \mu_{0}^2(\mu_{0}^2)=\frac{\lambda}{2}
    \frac{1}{\mom \sqrt{\mu_{0}^2 +4}} 
    F\left(\frac{\pi}{2}, \frac{2}{\sqrt{\mu^2+4}}\right).
\end{align}
Here, $F$ is the elliptic integral of the first kind
\begin{align}
    F\left(\alpha,m\right)=\int_{0}^{\alpha}\frac{1}{\sqrt{1-m^2 \sin(x)}}\diff x.
\end{align}
The counter-term is derived from the only divergent Feynman diagram of the theory in $(1\!+\!1)\mathrm{D}$, depicted in \cref{fig: Feynman-diagramm}.

The necessity for renormalizing the theory by subtractions comes from the choice of a real-space realization of the Hamiltonian. 
In contrast, in Fock space, the theory can be regularized by choosing the normal-ordered Hamiltonian~\cite{Glimm2012QuantumPhysicsFunctional}.

\section{Numerical methods}
In this section, we review the three numerical tools adopted to extract the local RDM of the given QMB Hamiltonians and provide the corresponding local basis: exact diagonalization (ED), k-side cluster mean-field (MF) theory, and tensor network (TN) methods.
\subsection{k-side MF theory}
\label{app_MF}
To formulate the $k$-site MF problem, we define the $k$-site MF state as
\begin{align}
    \ket{\psi(a,b)}
    \!=\!\left[\bigotimes_{-\infty}^{n-1} \ket{\psi (a)}\right]
    \!\otimes\! \ket{\psi(b)} 
    \!\otimes\!\left[\bigotimes_{n+k}^{\infty} {\psi(a)}\right]
    \!\in\! \mathcal{H}^{\infty}_{d},
\end{align}
where $\mathcal{H}^{\infty}_d$ denotes an infinite lattice system with local Hilbert space dimension $d$ at each site. The states $\ket{\psi(a)}$ and $\ket{\psi(b)}$ are normalized vectors defined on the $k$-site subsystem $\mathcal{H}_d^{\otimes k} \subset \mathcal{H}^{\infty}_d$. 
Here, the parameters $a$ and $b$ indicate that the corresponding states are variational.

The $k-$side MF estimation for the many-body ground-state minimizes then $\mathrm{min}_{\psi} \braket{\psi|H|\psi} $, under the constraint that
$\ket{\psi}$ is normalized. 
To achieve that, we formulate the Lagrange function
\begin{align}
    L=\langle \psi |\Ham |\psi\rangle-\lambda\left(\langle \psi|\psi\rangle-1\right)
    \label{eq: opt_prob}
\end{align}
and impose 
\begin{align}
    \frac{\partial L}{\partial \bra{\psi (b)}}=0,
    \label{eq: min_varphi_b}
\end{align}
which is minimized if $\ket{\psi(b)}$ is an eigenvector of an effective Hamiltonian (for $k\!=\!2$, see \cref{fig: E_eff_k=2}).
We start the algorithm by sampling a random state $\ket{\psi(a)}\!\in\!\mathcal{H}_{d}^{\otimes k}$, calculate $\ket{\psi(b)}$, then $\ket{\psi(a)}\!\leftarrow\!\ket{\psi(b)}$ and repeat to minimize with respect to $\ket{\psi(b)}$ until convergence in consecutive ground state energies $|E_{k}\!-\!E_{k-1}|\!<\!\epsilon$ for a chosen value $\epsilon >0$. 

\begin{figure}[ht]
    \centering
    \includegraphics[width=0.9\columnwidth]{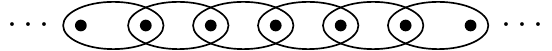}
    \caption{MF state for $k=2$}  
    \label{eq: k-2_mean-field}
\end{figure}

\begin{figure}[ht]
    \centering
    \includegraphics[width=0.9\columnwidth]{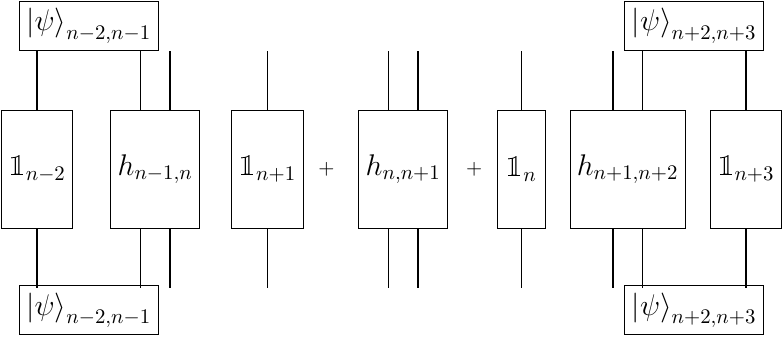}
    \caption{Effective operator to minimize for a two-sided MF state and Hamiltonian $H=\sum_{i}h_{i, i+1}$ with nearest neighbour two-body operators $h_{i, i+1}$.}
    \label{fig: E_eff_k=2}
\end{figure}

An example of a 2-side MF state is depicted in \cref{eq: k-2_mean-field}.
Taking a two-sided MF state as an ansatz, the effective Hamiltonian with nearest-neighbour interaction can then be depicted as in \cref{fig: E_eff_k=2}.

\subsection{Tensor Network}
\label{appendix: tensor_network}
To study the ground-state properties of the models, we employ a TN technique, such as matrix product states (MPS)~\cite{Fannes1992FinitelyCorrelatedStates-1} and tree tensor network (TTN)~\cite{Shi2006ClassicalSimulationQuantum-1}\ states. 
A general quantum many-body (QMB) state is defined on a lattice $\Lambda$ with $N = |\Lambda|$ sites. 
A state $\ket{\psi}$ of this quantum system lives in the Hilbert space
\begin{math}
    \mathcal{H} = \bigotimes_{i \in \Lambda} \mathcal{H}_{i},
\end{math}
where $\mathcal{H}_{i}$ denotes the local Hilbert space at site $i$, each of dimension $d$. 
The full state $\ket{\psi}$ can be expanded as
\begin{equation}
    \ket{\psi} = \sum_{i_{1}, \ldots, i_{N}} \psi_{i_{1}, \ldots, i_{N}} \ket{i_{1}, \ldots, i_{N}},
\end{equation}
where $\{\ket{i}_{j}\}_{i=1}^{d}$ is an orthonormal basis of $\mathcal{H}_{j}$. 
A complete description of the quantum state $\ket{\psi}$ thus requires tracking $d^N$ complex coefficients, which becomes computationally unfeasible for large $N$.

However, many low-energy QMB states exhibit limited entanglement. 
This behaviour is captured by the entanglement area law, which states that the entanglement entropy of a subsystem scales with the size of its boundary rather than its volume. 
When the area law holds, an efficient representation of $\ket{\psi}$ as an MPS exists~\cite{Eisert2010ColloquiumAreaLaws-1},
\begin{equation}
    \ket{\psi} = \sum_{i_1,\dots,i_{N}}
    \Tr\left[A_{1}^{i_{1}} \cdots A_{N}^{i_{N}}\right] 
    \ket{i_{1},\dots,i_{N}},
    \label{eq: MPS}
\end{equation}
where the number of parameters scales polynomially with the system size.
In~\cref{eq: MPS}, for each physical index $i_{j}$, the matrix $A_{j}^{i_{j}}$ has shape $D_{j-1} \times D_{j}$. In the case of open boundary conditions, the boundary matrices reduce to row and column vectors, such that $D_{1} = D_{N} = 1$.
In one-dimensional systems with gapped local Hamiltonians, where locality refers to interactions between neighbouring sites, the area law has been proven analytically. 
For higher-dimensional QMB systems, no general proof exists; however, extensive numerical evidence supports the validity of the area law under the assumption of local interactions \cite{Eisert2010ColloquiumAreaLaws-1, Masanes2009AreaLawEntropy-1, Kastoryano2019LocalityBoundaryImplies-1}.
To estimate the ground state and low-lying excitations, we apply the density-matrix renormalization group (DMRG) algorithm
~\cite{Schollwock2005DensitymatrixRenormalizationGroup, Verstraete2008MatrixProductStates-1}. 
All the TN simulations performed in this work have been obtained using the open-source TN library Quantum TEA \cite{Baccari2025QuantumTEAQtealeaves-1, Baccari2025QuantumTEAQredtea-1}.

\section{Dressed-site formulation of Hamiltonian Lattice Gauge Theories}
\label{app_dressed_site}

The Gauss operator for a compact Lie gauge group $\text{G}$ can be written as  
\begin{equation}
    \Op{G}[\vecsite]^{\nu}=\Op{Q}[\vecsite]^{\nu}+\Op{q}[\vecsite]^{\nu}
    +\sum_{\mu}
    \left[
    \Op{L}[\genlink]^{\nu}+\Op{R}[\vecsite-\vb{\mu},\vb{\mu}]^{\nu}
    \right],
    \label{eq: gauss_operator}
\end{equation}
where $\Op{L}[\genlink]^{\nu}$ and $\Op{R}[\genlink]^{\nu}$ are the independent generators of the left and right gauge transformations, located at the respective ends of the link $(\genlink)\in \Lambda$. The index $\nu$ labels the generators, with $\nu\in \{1,\dots,\dim(\text{G})\}$.  
In the Abelian $U(1)$ case, the left and right generators coincide and reduce to the electric field,  
\begin{math}
    \Op{E}[\genlink]\!=\!
    \Op{L}[\genlink]\!=\!
    \Op{R}[\genlink].
\end{math}  
In~\cref{eq: gauss_operator}, $\Op{q}[\vecsite]^{\nu}$ denotes the background charge, which is set to zero in the present work, while  
\begin{math}
    \Op{Q}[\vecsite]^{\nu}
    =\sum_{\alpha,\beta} 
    \Op{\psi}[\vecsite,\alpha][1] 
    \Op{T}[\alpha,\beta]^{\nu}
    \Op{\psi}[\vecsite,\beta]
\end{math}  
represents the matter charge at lattice site $\vecsite$, with $\{\Op{T}^{\nu}\}_{\nu=1}^{\dim(\text{G})}$ denoting the generators of the gauge group.

To locally enforce Gauss’s law in LGT, as discussed in~\cref{sec_LGTs}, we employ the so-called dressed-site formalism. 
This approach reduces a non-Abelian constraint to an Abelian one, as we will see in this section. 
Although dressed sites typically have larger local dimensions than those arising from other truncation schemes, this approach applies to all compact gauge groups. 
It preserves the locality of interactions, making the framework particularly well-suited for TN simulations.
First, the parallel transporter is decomposed into left and right degrees of freedom, referred to as left (L) and right (R) rishon operators. 
This is achieved by embedding the attached link Hilbert space as~\cite{Silvi2014LatticeGaugeTensor-1}
\begin{math}
    \ket{j m n} \;\to\; \ket{j m}_L \otimes \ket{j n}_R .
\end{math}
The parallel transporter then takes the form
\begin{equation}
  \Op{U} \to \sum_{k}
  \zeta^{L(k)\alpha}_{\vecsite,+\vb*{\mu}}\zeta^{R(k)\beta\,\dagger}_{\vecsite+\vb*{\mu},-\vb*{\mu}} \,
  \label{eq_from_U_to_rishons}
  \,.
\end{equation}
Physical configurations are those in which the left and right rishons are in the same irreducible representation. 
This construction thus yields an Abelian link symmetry on the TN level~\cite{Silvi2014LatticeGaugeTensor-1, Rigobello2023Hadrons1+1DHamiltonian-1}, regardless of whether the gauge symmetry of the theory is Abelian or non-Abelian.

Now, the local gauge generators act only on the matter site at $x$ and its $2D$ neighboring rishons in the lattice $\Lambda$. 
The matter site and the surrounding rishon degrees of freedom are fused into a composite site, and Gauss’s law becomes a constraint that reduces the dressed-site Hilbert space to the Hilbert space of physical configurations, the states transforming under the singlet
\begin{equation}
    \mathcal{H}_{dress}= \text{ker}\, \Op{G}[\vecsite]^{a} \subset 
    \mathcal{H}_{\text{matt}} \otimes \left(\mathcal{H}_{\text{rish}}\right)^{\otimes 2D}.
\end{equation}
The gauge-singlet basis states of $\mathcal{H}_{\text{dress}}$ are then expanded, the Clebsch–Gordan~\cite{Wigner2012GroupTheoryIts} decomposition in terms of matter and rishon bases.
\end{document}